# Rotary solitons in Bessel photonic lattices

Yaroslav V. Kartashov,[1,2] Victor A. Vysloukh,[3] Lluis Torner[1]

[1]*ICFO-Institut de Ciencies Fotoniques, and Department of Signal Theory and Communications, Universitat Politecnica de Catalunya, 08034, Barcelona, Spain*

[2]*Physics Department, M. V. Lomonosov Moscow State University, 119899, Moscow, Russia*

[3]*Departamento de Fisica y Matematicas, Universidad de las Americas – Puebla, Santa Catarina Martir, 72820, Puebla, Mexico*

We introduce solitons supported by Bessel photonic lattices in cubic nonlinear media. We show that the cylindrical geometry of the lattice, with several concentric rings, affords unique soliton properties and dynamics. In particular, besides the lowest-order solitons trapped in the center of the lattice, we find soliton families trapped at different lattice rings. Such solitons can be set into controlled rotation inside each ring, thus featuring novel types of in-ring and inter-ring soliton interactions.

Propagation of optical radiation in media whose guiding properties vary in transverse directions exhibits a wealth of practically interesting phenomena which find applications in different branches of physics, including waves in molecular chains [1], trapped Bose-Einstein condensates [2], or solids [3]. In particular, transverse modulation of refractive index can profoundly affect the properties of spatial optical solitons formed in nonlinear medium through the competition of diffraction and nonlinear self-action effects. In the case of periodic refractive index modulation discrete solitons are known to be formed in the evanescently coupled guiding sites of the array [4]. The interest in propagation of such solitons is dictated mainly by their rich potential for all-optical switching, power and angle-controlled steering [5]. Recently it was predicted that lattices constituted by continuous nonlinear media with an imprinted harmonic modulation of the refractive index offer a number of opportunities for all-optical manipulation of light signals as well [6-8], because such lattices may operate in both weak- and strong-coupling regimes [9] depending on the depth of refractive index modulation. In landmark recent experiments [10,11], it was demonstrated that periodic waveguide lattices with flexibly controlled refractive index modulation depth and waveguide separation can be formed optically, in particular in photorefractive media. Besides lowest-order single solitons, harmonic lattices support optical vortices [12,13] and soliton trains [14].

To the date main efforts were devoted to analysis of spatial solitons supported by square, or honeycomb two-dimensional lattices that can be induced optically with several interfering plane waves. The properties of such solitons depend crucially on the lattice symmetry. Thus, the periodicity of harmonic lattices dictates the structure of the existence domain of lowest-order solitons as well as the symmetry of field distribution for lattice vortices. Therefore, the fascinating question that arises is whether optical lattices with different types of symmetry offer new opportunities for soliton existence, managing and control.

In this Letter we address the basic properties and linear stability of optical solitons supported by radial-symmetric Bessel lattices in cubic nonlinear media. Such lattices are especially interesting because of the suppressed diffraction for the lattice-inducing field, similarly to the case of harmonic lattices. Besides the expected guidance in the central core of the Bessel lattice, we found that localized bell-shaped solitons can be supported by different lattice rings. We also found that such solitons can rotate inside these rings without energy radiation. One of the central results we put forward here is the possibility of controlling the interactions of solitons located within the same or in separated rings by inducing their rotary motions.

We address the propagation of optical radiation along the $z$ axis of a bulk cubic medium with a transverse modulation of linear refractive index, described by the nonlinear Schrödinger equation for the dimensionless complex field amplitude $q$:

$$i\frac{\partial q}{\partial \xi} = -\frac{1}{2}\left(\frac{\partial^2 q}{\partial \eta^2} + \frac{\partial^2 q}{\partial \zeta^2}\right) + \sigma q|q|^2 - pR(\eta,\zeta)q. \quad (1)$$

Here the longitudinal $\xi$ and transverse $\eta, \zeta$ coordinates are scaled to the diffraction length and to the input beam width, respectively, $\sigma = \mp 1$ for focusing/defocusing nonlinearity. The guiding parameter $p$ is proportional to the refractive index modulation depth; the function $R(\eta,\zeta)=J_0[(2b_{\text{lin}})^{1/2}(\eta^2+\zeta^2)^{1/2}]$ stands for the transverse profile of refractive index; the parameter $b_{\text{lin}}$ is related to the radii of rings in a zero-order Bessel lattice. We assume that the depth of the refractive index modulation is small compared with the unperturbed index and is of the order of the nonlinear contribution to refractive index. Here we address the case of the simplest zero-order Bessel lattice. Eq. (1) admits some conserved quantities including the total energy flow $U=\int_{-\infty}^{\infty}\int_{-\infty}^{\infty}|q|^2\,d\eta d\zeta$ .

The function $q(\eta,\zeta,\xi)=J_0[(2b_{\text{lin}})^{1/2}(\eta^2+\zeta^2)^{1/2}]\exp(-ib_{\text{lin}}\xi)$ is an exact solution of the linear homogeneous Eq. (1) at $\sigma=p=0$ describing a nondiffracting laser beam [15]. Rigorously such beams extent to the transverse infinite, but accurate approximations can be generated experimentally in a number of ways. Known methods include simple techniques like annular slits in the focal plane of a lens and conical axicons, as well as more elaborated interferometric and holographic techniques (see Refs [16] and references quoted therein). It is worth stressing that holographic techniques allow producing not only single Bessel beams, but also more complicated families of nondiffracting beams. The Bessel photonic lattices addressed in this paper can thus be imprinted, e.g., in photorefractive crystals, using the techniques recently reported for the generation of harmonic patterns by incoherent vectorial interactions [10-13]. Because of the diffractionless nature of the Bessel beams, they are to be launched collinearly along the material with a polarization orthogonal to the soliton beams.

First we address the properties of lowest-order solitons with the radial symmetry supported by a central guiding core of Bessel lattice. We search for solution of Eq. (1) in the form $q(\eta,\zeta,\xi)=w(r)\exp(ib\xi)$, where $r^2=\eta^2+\zeta^2$, $b$ is the propagation constant, and $w(r)$ is a real function. The resulting ordinary differential equation for the function $w(r)$ was solved with a standard relaxation algorithm. Lattice soliton families are defined by the propagation constant $b$, lattice and guiding parameters $b_{\text{lin}}$ and $p$. Since one can use scaling transformation $q(\eta,\zeta,\xi,p)\rightarrow\chi q(\chi\eta,\chi\zeta,\chi^2\xi,\chi^2 p)$ to obtain various families of lattice solitons from a given family, we selected the transverse scale in such a way that $b_{\text{lin}}=10$ and vary $b$ and $p$. Below we discuss focusing nonlinear media with $\sigma=-1$. The properties of lowest order solitons whose intensity maximum coincides with the axis of the lattice are summarized in Fig. 1. At low energy flows solitons are wide and cover several lattice rings, while at high energies they are narrow and concentrate mostly within the core of the lattice (Fig. 1(b)). At small values of guiding parameter, the energy flow is a non-monotonic function of the propagation constant, while above the critical level $p \gtrsim 3.91$, the energy flow increases monotonically with growth of propagation constant (Fig. 1(a)). There exists a lower cut-off $b_{\text{co}}$ on propagation constant that increases with growth of refractive index modulation depth (Fig. 1(c)). At $b \to b_{\text{co}}$ the soliton energy flow goes to zero.

To elucidate the linear azimuthal modulational stability of the above soliton families we searched for perturbed solution of Eq. (1) with the form $q(\eta,\zeta,\xi) = [w(r) + u(r,\xi)\exp(in\phi) + v^*(r,\xi)\exp(-in\phi)]\exp(ib\xi)$, where the perturbation components $u, v$ grow with the complex rate $\delta$ upon propagation, $\phi$ is an azimuthal angle, and $n = 0,1,2...$ is the azimuthal index. Linearization of Eq. (1) around the stationary solution $w(r)$ yields the eigenvalue problem:

$$\begin{aligned} i\delta u &= -\frac{1}{2}\left(\frac{d^2}{dr^2} + \frac{1}{r}\frac{d}{dr} - \frac{n^2}{r^2}\right)u + bu + \sigma w^2(v + 2u) - pRu, \\ -i\delta v &= -\frac{1}{2}\left(\frac{d^2}{dr^2} + \frac{1}{r}\frac{d}{dr} - \frac{n^2}{r^2}\right)v + bv + \sigma w^2(u + 2v) - pRv, \end{aligned} \tag{2}$$

which was solved numerically. We have found that (in agreement with the Vakhitov-Kolokolov stability criterion) lowest-order solitons supported by Bessel lattices are stable provided that the condition $dU/db > 0$ is satisfied. Notice that at $p \lesssim 3.91$ there exists a narrow band of propagation constants near cut-off where $dU/db \le 0$ and thus the corresponding solitons are unstable (see inset in Fig. 1(c) for the borders of instability band and Fig. 1(d) for dependence of the growth rate with the propagation constant). When the guiding parameter reaches a critical value, we found that the solitons become stable in the entire domain of their existence.

A fascinating example of localized self-sustained light structure supported by Bessel lattices is given by solitons trapped in different rings of the lattice. Solitons in the ring can be formed due to the local radial maximum of the refractive index. We searched for profiles of such solitons in the form $q(\eta,\zeta,\xi) = w(\eta,\zeta)\exp(ib\xi)$. An example of a soliton trapped in the first ring of a $J_0$ lattice is shown in Fig. 2(a) (lattice rings supporting solitons are shown by circles). The energy flow is a non-monotonic function of the propagation constant (Fig. 2(b)). At low energy flows soliton becomes spatially extended and covers several neighboring lattice rings. The cut-off on propagation constant $b_{co}$ increases monotonically with growth of the guiding parameter.

To elucidate the stability of solitons trapped in the rings, we performed extensive set of simulations of Eq. (1) with an input conditions $q(\eta,\zeta,\xi=0)=w(\eta,\zeta)[1+\rho(\eta,\zeta)]$, where $\rho(\eta,\zeta)$ describes a broadband random perturbation. Simulations revealed that solitons suffer from exponential instabilities close to cut-off and become completely stable above a certain energy threshold. The corresponding solitons are relatively narrow and their overlap with neighboring rings diminishes. The threshold energy flow that is necessary for stabilization decreases with growth of refractive index modulation depth.

One of central results of this Letter is the possibility to induce rotary motion of solitons trapped inside the ring (Figs. 2(c) and 2(d)). This can be done either by launching a laser beam tangentially to the lattice rings or by imposing the phase twist $\exp(i\nu\phi)$ on the stationary soliton solutions described above. The important effect is that upon rotation inside the input ring soliton does not radiate, and thus survives for thousands of diffraction lengths even in the presence of broadband input noise. If the imposed phase twist $\nu$ (or launching angle) exceeds a critical value, the soliton leaves the ring where it was initially located and moves across the lattice until it complete decay under the influence of radiative losses. The larger the refractive index modulation depth the higher the accessible value of the phase twist, and hence the higher the soliton angular frequency. The possibility of excitation of rotary motion of solitons inside Bessel lattice might find direct applications in future soliton circuits. Among other options is the possibility of controlling the output soliton angle and soliton position.

We have found a rich variety of interaction scenarios for *collective* motion of solitons inside the lattice rings. Some illustrative examples of such dynamical structures of soliton pairs are presented in Fig. 3. Usually we set one of identical solitons (upper one) into rotation by superimposing a phase twist, while the second soliton was initially at rest. For low $\nu$ values the interaction is highly sensitive to phase difference between solitons. Out-of-phase solitons tend to repel each other and as a result a steadily rotating soliton pair with fixed separation between components is formed despite the fact that one of the solitons was not set initially into motion (column (a)). This indicates the existence of rotating stable complexes that consist of solitons with alternating phases located within one ring. In-phase solitons fuse upon collision into one soliton of higher intensity rotating with lower angular frequency than the initial one (column (b)). When the phase twist imposed on the upper soliton is high enough we found that the two interacting solitons always fuse upon collision.

A remarkable scenario was encountered upon the interaction of out-of-phase solitons trapped in different lattice rings (column (c)). One finds that the soliton in the internal ring is able to reverse the rotation direction upon the first interaction. This results in rotation of solitons located in different rings in opposite directions; the consecutive interactions (or collisions) of the solitons is then accompanied by reversing the rotation direction. Such lattice soliton interactions might be the basis of a new class of soliton-based blocking and routing schemes [5].

Finally we illustrate an example of a complex *static* soliton structure supported by the Bessel lattice that can be viewed as a combination of out-of-phase solitons supported by the first ring and the core (Fig. 4(a)). Such structures are somehow analogous to twisted (dipole-mode) soliton in harmonic lattices. Their energy flow becomes a monotonically growing function of propagation constant for high enough values of guiding parameter (Fig. 4(b)). Numerical simulations showed that such twisted solitons (Fig. 4(a)) suffer from oscillatory instabilities near their cut-off and become stable above an energy threshold. The structure of such twisted solitons suggests the possibility to pack several solitons (Fig. 4(d)) with appropriately engineered phases into a compact composite and then to extract solitons from it. Notice that these twisted solitons can rotate around the lattice rings upon propagation.

To conclude, we revealed that the cylindrical geometry of Bessel photonic lattices, with several concentric radial rings, affords a wealth of new soliton features, both stationary and dynamical. In particular, we found soliton families trapped at different lattice rings, which can be set into controlled rotation inside each ring, thus featuring unique types of soliton interactions. Here we considered cubic nonlinear media, but the concept can be extended to all relevant physical settings, including other types of optical nonlinearities and Bose-Einstein condensates.

This work has been partially supported by the Generalitat de Catalunya and by the Spanish Government through grant BFM2002-2861.

## Figure captions

Figure 1. (a) Energy flow versus propagation constant. (b) Lattice soliton profiles corresponding to points marked by circles in (a) and structure of Bessel lattice (inset). (c) Lower cutoff and critical values of propagation constant for stabilization (inset) versus guiding parameter. (d) Perturbation growth rate versus propagation constant for different guiding parameters.

Figure 2. (a) Profile of soliton trapped in the first lattice ring at $p = 5$ corresponding to point marked by circle in the dispersion diagram (b). Snapshot images showing soliton rotation in the (c) first and (d) second rings of Bessel lattice. Solitons correspond to propagation constant $b = 3$ at $p = 5$. Parameter $\nu = 1$. Step in propagation distance between different snapshots is $\delta\xi = 2.5$ in (c) and $\delta\xi = 5$ in (d).

Figure 3. (a) Interaction of out-of-phase solitons in the first ring of the lattice. One of the solitons is set into motion at the entrance of the medium by imposing the phase twist $\nu = 0.1$. (b) The same as in column (a) but for in-phase solitons. (c) Interaction of out-of-phase solitons located in the first and second lattice rings. The soliton located in the first ring is set into motion by imposing the phase twist $\nu = 0.2$. All solitons correspond to the propagation constant $b = 5$ at $p = 5$. Soliton angular rotation directions are depicted by arrows.

Figure 4. (a) Profile of the first twisted soliton at $b = 2$, $p = 6$. (b) Energy flow versus propagation constant for different guiding parameters. (c) Propagation constant cutoff versus guiding parameter. (d) Profile of second twisted soliton at $b = 2$, $p = 6$.

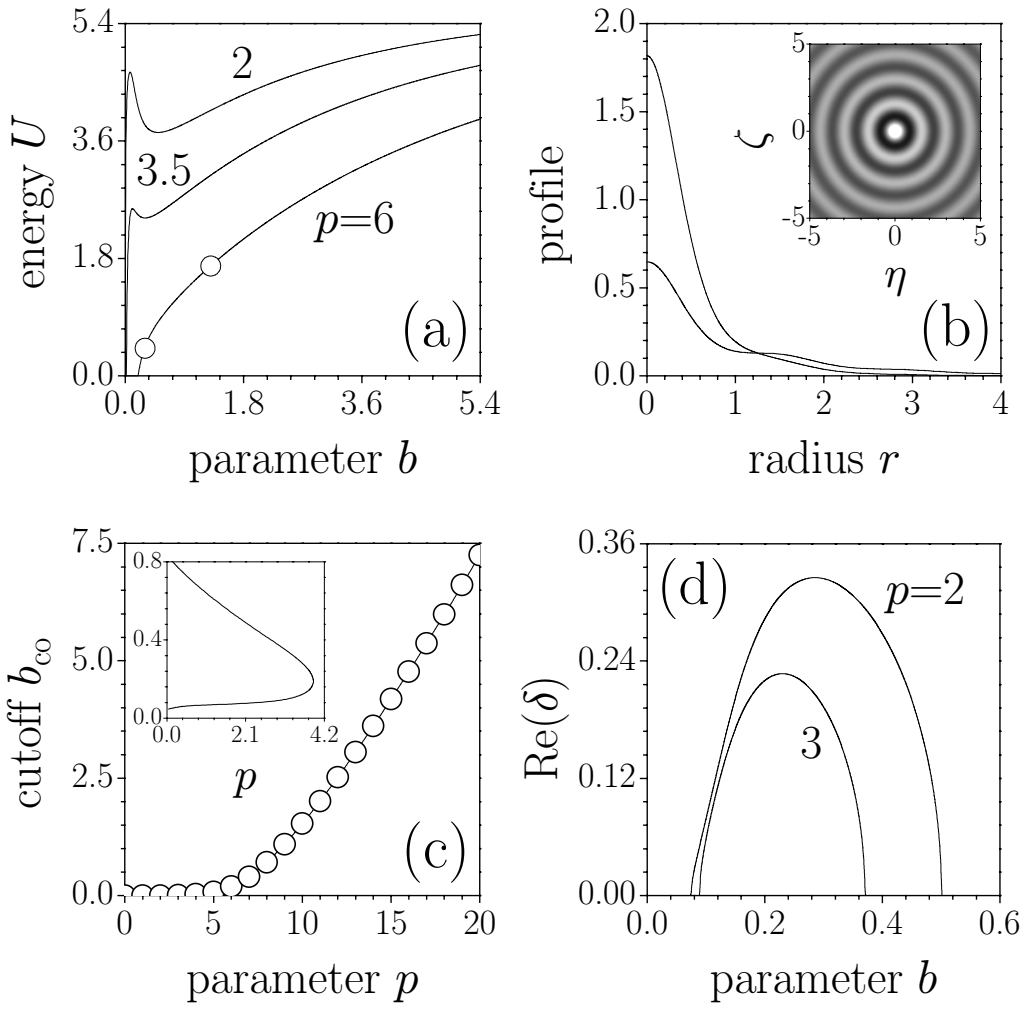

Figure 1. (a) Energy flow versus propagation constant. (b) Lattice soliton profiles corresponding to points marked by circles in (a) and structure of Bessel lattice (inset). (c) Lower cutoff and critical values of propagation constant for stabilization (inset) versus guiding parameter. (d) Perturbation growth rate versus propagation constant for different guiding parameters.

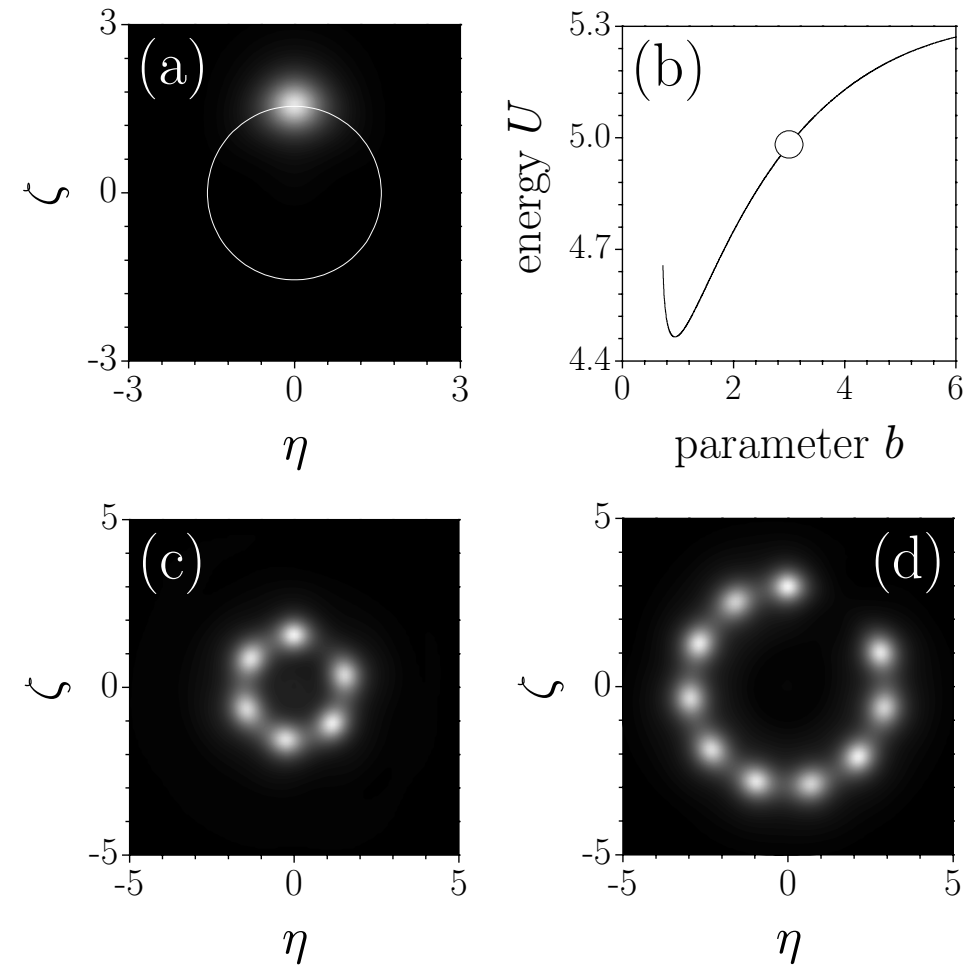

Figure 2. (a) Profile of soliton trapped in the first lattice ring at $p = 5$ corresponding to point marked by circle in dispersion diagram (b). Snapshot images showing soliton rotation in the (c) first and (d) second rings of Bessel lattice. Solitons correspond to propagation constant $b = 3$ at $p = 5$. Parameter $\nu = 1$. Step in propagation distance between different images is $\delta\xi = 2.5$ in (c) and $\delta\xi = 5$ in (d).

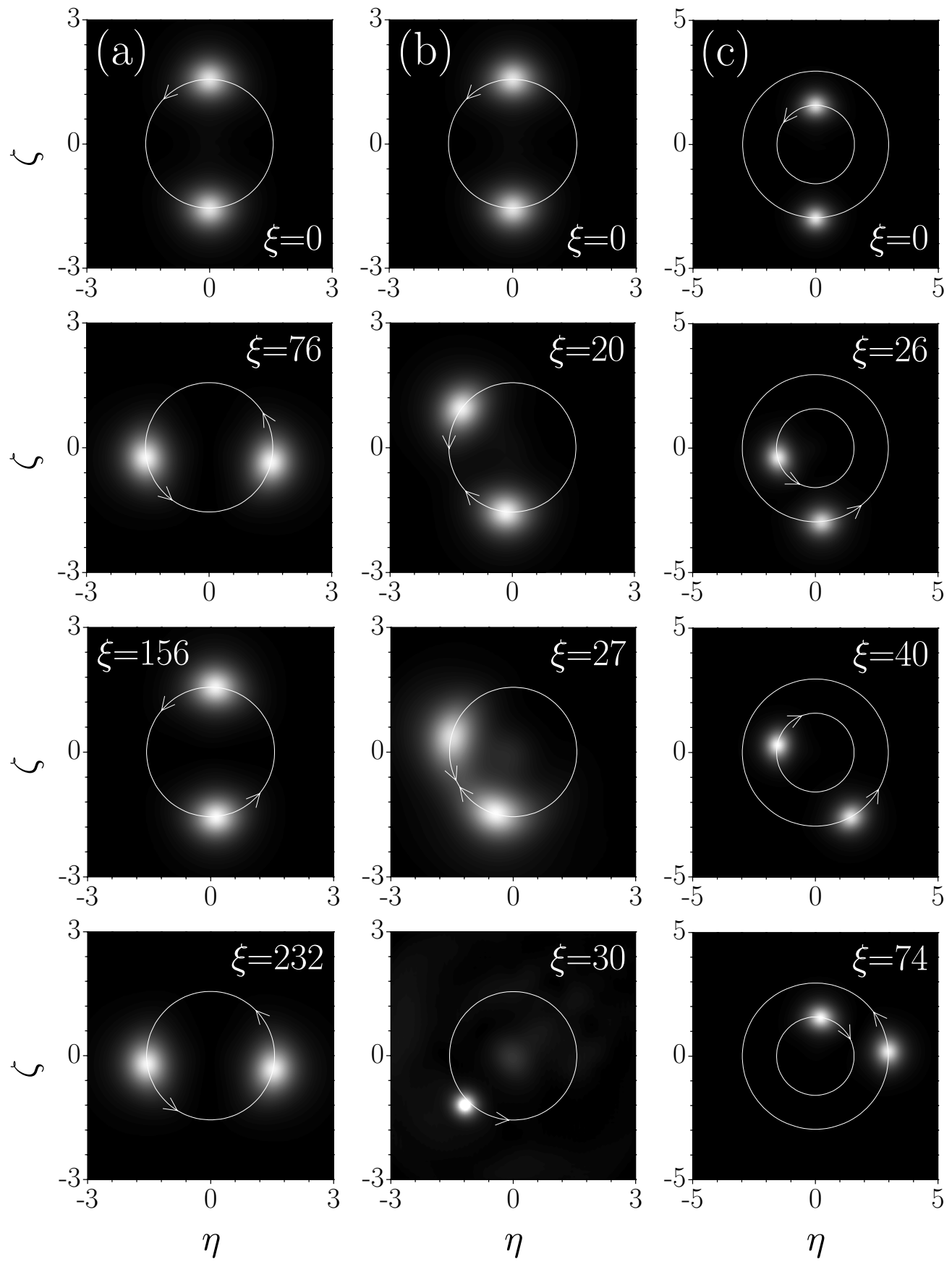

Figure 3. (a) Interaction of out-of-phase solitons in the first ring of the lattice. One of the solitons is set into motion at the entrance of the medium by imposing the phase twist $\nu = 0.1$. (b) The same as in column (a) but for in-phase solitons. (c) Interaction of out-of-phase solitons located in the first and second lattice rings. The soliton located in the first ring is set into motion by imposing the phase twist $\nu = 0.2$. All solitons correspond to the propagation constant $b = 5$ at $p = 5$. Soliton angular rotation directions are depicted by arrows.

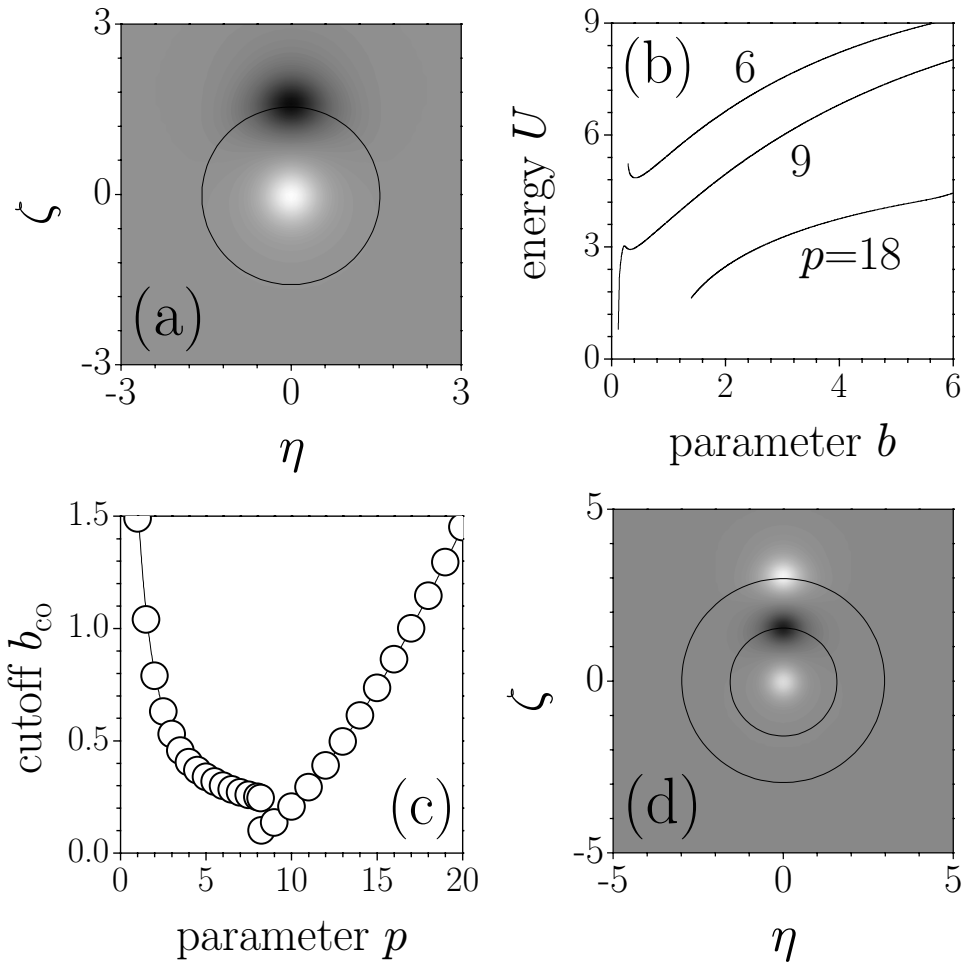

Figure 4. (a) Profile of first twisted soliton at $b = 2$, $p = 6$. (b) Energy flow versus propagation constant for different guiding parameters. (c) Propagation constant cutoff versus guiding parameter. (d) Profile of second twisted soliton at $b = 2$, $p = 6$.